\documentclass{aastex631}

\usepackage{color}
\definecolor{bbsalmon}{rgb}{1.0, 0.47, 0.42}

\begin{document}

\title{Tracking the Evolution of Supermassive Black Hole Binaries with Pulsar Terms}

\author[0000-0003-0909-5563]{Bence B\'{e}csy}
\affiliation{Institute for Gravitational Wave Astronomy and School of Physics and Astronomy, University of Birmingham, Edgbaston, Birmingham B15 2TT, UK}

\author[0000-0002-1530-9778]{Magdalena Siwek}
\affiliation{Columbia University, Pupin Hall, 538 W 120th St, New York, NY 10027, USA}
\affiliation{New York University, Center for Cosmology and Particle Physics, 726 Broadway, New York, NY 10003, USA}

\begin{abstract}
Pulsar timing arrays will soon be able to resolve the loudest supermassive black hole binaries contributing to the stochastic gravitational wave background at nanohertz frequencies. While the few decades of observations are typically not enough to see these binaries evolve in frequency, the so-called pulsar terms offer a glimpse into the past and can help us map out the frequency evolution of these binaries over tens of thousands of years. Measuring this long-term frequency evolution can allow us to look for deviations from a vacuum binary evolution due to various environmental effects. We develop a Fisher-matrix approach to calculate how precisely the frequency can be tracked over time in this manner, and demonstrate its use on a realistic near-future pulsar timing array with a 30-year baseline. We find that the environmental effects predicted by 2-dimensional hydrodynamic simulations of circumbinary disks are unlikely to be detectable in the circular binaries we considered, however, in some cases they might result in biased chirp mass recovery if ignored. We also demonstrate how this approach allows one to test the orbital evolution predicted by general relativity in a purely gravitational-wave driven inspiral, and to put upper limits on additional generic environmental effects described by a power-law hardening model.
\end{abstract}


\section{Introduction} \label{sec:intro}

Supermassive black holes (SMBHs) are the most massive black holes in the universe, which reside at the centers of the most massive galaxies. Following the merger of such galaxies, the black holes can form an SMBH binary (SMBHB), which can emit gravitational waves (GWs) in the nano-millihertz frequency range. In particular, the most massive of these binaries will be predominantly detectable at nanohertz frequencies, where pulsar timing array (PTA) experiments are sensitive. The most recent PTA datasets \citep{nanograv_15yr_data, epta_dr2_data, ppta_dr3_data, MPTA_4p5yr_data} show evidence for the presence of a stochastic GW background (GWB) \citep{nanograv_15yr_gwb, epta_dr2_gwb, ppta_dr3_gwb, cpta_dr1_gwb, MPTA_4p5yr_gwb}. The most likely source of such a GWB is the collective signal from a large ensemble of SMBHBs \citep{nanograv_15yr_astro, epta_dr2_astro}. If that is the case, we also expect to be able to resolve the loudest SMBHBs individually within the next decade or so \citep{SVV2009, RosadoExpectedProperties, ChiaraLandscape,Luke_single_source,RealisticDetection1,Emiko2024,Emiko2025}.

There are several astrophysical processes that can drive the evolution of SMBHBs, which are crucial in order to reach sub-parsec separations where GW emission can effectively drive the binary to merger \citep{BegelmanBlandfordRees1980, MilosavljevicMerritt2003}. These can include interactions with a gaseous environment \citep{KocsisSesana2011} and individual stars in the nuclear star cluster \citep{Khan_scattering2013}, or more exotic effects such as dynamical friction inside a dark matter density spike \citep{DM_finalparsec_2024}. These processes can be observationally tested if they keep contributing to the evolution of the SMBHB while it is already in the observable frequency band for GW detectors. In particular, this can cause a turnover in the GWB spectrum at the lowest frequencies observable by PTAs \citep{KocsisSesana2011, Sampson2015}. Although the effects of an environment on a GWB have been extensively studied, their influence on resolvable binaries largely remained unexplored in the PTA regime. It has been shown however, that in the millihertz regime, environmental effects can leave interesting observable signatures in the evolution of binaries (particularly extreme mass ratio inspirals, EMRIs), which can become detectable with the LISA mission \citep{speri_emri_environmental_effects,pippa_lisa_environmental_effects}.

SMBHBs in the PTA band are often modeled with a simple Newtonian chirp (0-PN) and assumed to be on a circular orbit (see e.g.~\citealt{NeilCWMethods, Lee_et_al_CW_methods, JustinCWMethods, Steve_accelerated_CW, QuickCW, FurgeHullam, AidenCW}), although analyses that allow for eccentricity are increasingly common \citep{steve_eccbin,abhimanyu_eccbin_2020,abhimanyu_eccbin_2022,nanograv_12p5yr_eccbin,ManziniBabak2026I,ManziniBabak2026II}. The simple treatment of binary evolution is often well justified, given that these systems are typically far away from merger and show little to no frequency evolution within the few decades of observations. However, PTAs are sensitive not only to local GWs (so-called Earth terms), but also to the GWs at the location of the pulsars. These pulsar terms probe the GW waveform from an SMBHB thousands (or tens of thousands) of years in the past. This opens the possibility of tracking the long-term frequency evolution of the binary, which can reveal signatures of various environments affecting the binary. \citet{Chiara_CW_evolution_tracking} explored how higher PN and spin effects can result in an accumulated phase difference between the pulsar and Earth terms. Here we extend this idea, and investigate how well environmental effects could be measured. We also improve on the methodology by introducing a Fisher formalism approach that can estimate a PTA's ability to track the time evolution of the GW frequency.

The rest of the paper is organized as follows. In Section \ref{sec:methods} we describe our Fisher formalism and the models we use for environmental effects. In Section \ref{sec:results} we apply our methods to a near future PTA setup and explore the detectability of various environments. Finally, we offer concluding remarks and discuss future directions in Section \ref{sec:conclusions}.

\section{Methods} \label{sec:methods}

Our goal is to assess how well the frequency evolution of SMBHBs can be tracked by PTA observations. To do so, we employ a Fisher formalism to predict how well each pulsar term frequency can be measured (Section \ref{ssec:fisher}). These act as frequency measurements in the past, where the error on the look-back time is determined by the measurement precision of pulsar distances. Comparing such measured frequencies and times with the frequency evolution predicted for a binary evolving under the influence of a gaseous environment (Section \ref{ssec:gas_drag}) or a generic power-law environment (Section \ref{ssec:generic_env}) allows us to constrain such environments (Section \ref{ssec:freq_evol_fit}). The code implementing these methods, together with example notebooks and scripts reproducing the main figures, is publicly available on GitHub \citep{cwcbd_github}.

\subsection{Fisher matrix formalism} \label{ssec:fisher}

The most reliable way to estimate the accuracy with which we can measure the frequency evolution of SMBHBs would be to carry out a full Bayesian analysis. This is however, prohibitively expensive to run, and would also require considerable development time to incorporate environmental effects into existing Bayesian analysis software. Instead, we opt for a Fisher matrix formalism, which provides a good approximation of the full results in the high signal-to-noise ratio (SNR) limit. Note that while this is an approximation, it is a more reliable approximation than calculating the phase or frequency shift introduced by an effect (as used e.g.~in \citealt{Chiara_CW_evolution_tracking}) and comparing that to some minimum measurable value. As we will see, one can often measure the frequency much better than the width of a ``frequency bin'', i.e.~the inverse of the observation time, as is often assumed in heuristic arguments of frequency evolution detectability.

A Fisher matrix treatment has previously been used on resolvable SMBHB signals in PTAs by \citet{tingting_cw_precision2023}. However, they only considered the Earth term, and focused on the measurement precision of the Earth term amplitude and frequency. Instead, we are interested in the measurement precision of both Earth and pulsar term frequencies, so our formalism will be different.

Our signal model has the following form:
\begin{equation}
    r_i = A_{\rm E} F_i \sin (2\pi f_{\rm E} t - \phi_{\rm E}) - (f_{\rm E}/f_i)^{1/3}A_{\rm E} F_i \sin (2\pi f_i t - \phi_i),
\end{equation}
where $A_{\rm E}$ is the Earth-term timing-residual amplitude, $F_i$ is the magnitude of the antenna patterns for the $i$th pulsar, $f_{\rm E}$ and $\phi_{\rm E}$ are the Earth-term frequency and phase, and $f_i$ and $\phi_i$ are the $i$th pulsar-term frequency and phase. Note that unlike the usual SMBHB models that connect the Earth and pulsar term frequencies via an evolution model, our model is generic, and parametrizes each pulsar term phase and frequency independently. So the model is described by 3+2$N_{\rm PSR}$ parameters $\theta = [A_{\rm E}, f_{\rm E}, \phi_{\rm E}, f_1, \phi_1, ..., f_{N_{\rm PSR}}, \phi_{N_{\rm PSR}}]$. Then the Fisher matrix can be written as:
\begin{equation}
    \mathcal{F}_{ij} = \sum_{k=1}^{N_{\rm PSR}} \frac{1}{\bar{\sigma}_k^2} \sum_{l=1}^{N_{\rm obs, k}} \frac{\partial r_{k,l}}{\partial \theta_i} \frac{\partial r_{k,l}}{\partial \theta_j},
\end{equation}
where $\bar{\sigma}_k$ is the average TOA error for the $k$th pulsar\footnote{Note that for simplicity we assumed uniform TOA errors for all observations of a given pulsar, but one could easily extend this formalism to include different errors for each observation by simply moving $\bar{\sigma}_i$ into the sum over observations and giving it one more index, or even to fully correlated errors by introducing the covariance matrix.}. Note that many terms in this sum are zero, as partial derivatives of a given pulsar's signal w.r.t.~a pulsar frequency or phase of another pulsar are zero. We use similar stabilization tricks as in \citet{tingting_cw_precision2023} to help with the inversion of the Fisher matrix. Once the inverse is obtained, the measurement error of $\theta_i$ is simply given by the square root of the corresponding diagonal element:
\begin{equation}
    \sigma_i = \sqrt{(\mathcal{F}^{-1})_{ii}}.
    \label{eq:fisher}
\end{equation}

We can also calculate the SNR of a given signal by:
\begin{equation}
    {\rm SNR} = \sqrt{\sum_{k=1}^{N_{\rm PSR}} \frac{1}{\bar{\sigma}_k^2} \sum_{l=1}^{N_{\rm obs, k}} r_{k,l}^2}.
\end{equation}

Thus, appropriate elements of $(\mathcal{F}^{-1})_{ii}$ give us the measurement precision of $f_{\rm E}$ and $f_i$. Importantly, these will correspond to the GW frequency of the SMBHB sampled at different times. Denoting the time at which the Earth term is measured as zero, a given pulsar term will correspond to the GW frequency at time:
\begin{equation}
    t_i = (\cos \beta_i-1)d_{\rm PSR, i}/c,
    \label{eq:psr_time}
\end{equation}
where $\beta_i$ is the angle between the pulsar's and the SMBHB's sky location, and $d_{\rm PSR, i}$ is the distance between the Earth and the pulsar. Note that this time is always negative, so pulsar terms probe the GW frequency in the past compared to the Earth term. We can also see that the time delay is zero if the source is right behind the pulsar, and is maximal if they are at antipodal sky locations. However, pulsars with large $\beta_i$ will see a weaker signal due to the antenna pattern function's $F_i \propto (1+\cos \beta_i)$ dependence. For typical pulsar distances of $\mathcal{O}(1 \ \rm kpc)$, the pulsar terms probe the GW frequency a few thousand years in the past. This can in principle be extended if the GW signal can be detected in further away pulsars \citep{Kathrin+2026_distant_psrs}, e.g.~pulsars in the LMC or SMC would probe the GW frequency a few hundred thousand years in the past.

\begin{figure}[htbp!]
    \centering
    \includegraphics[width = 0.6\columnwidth]{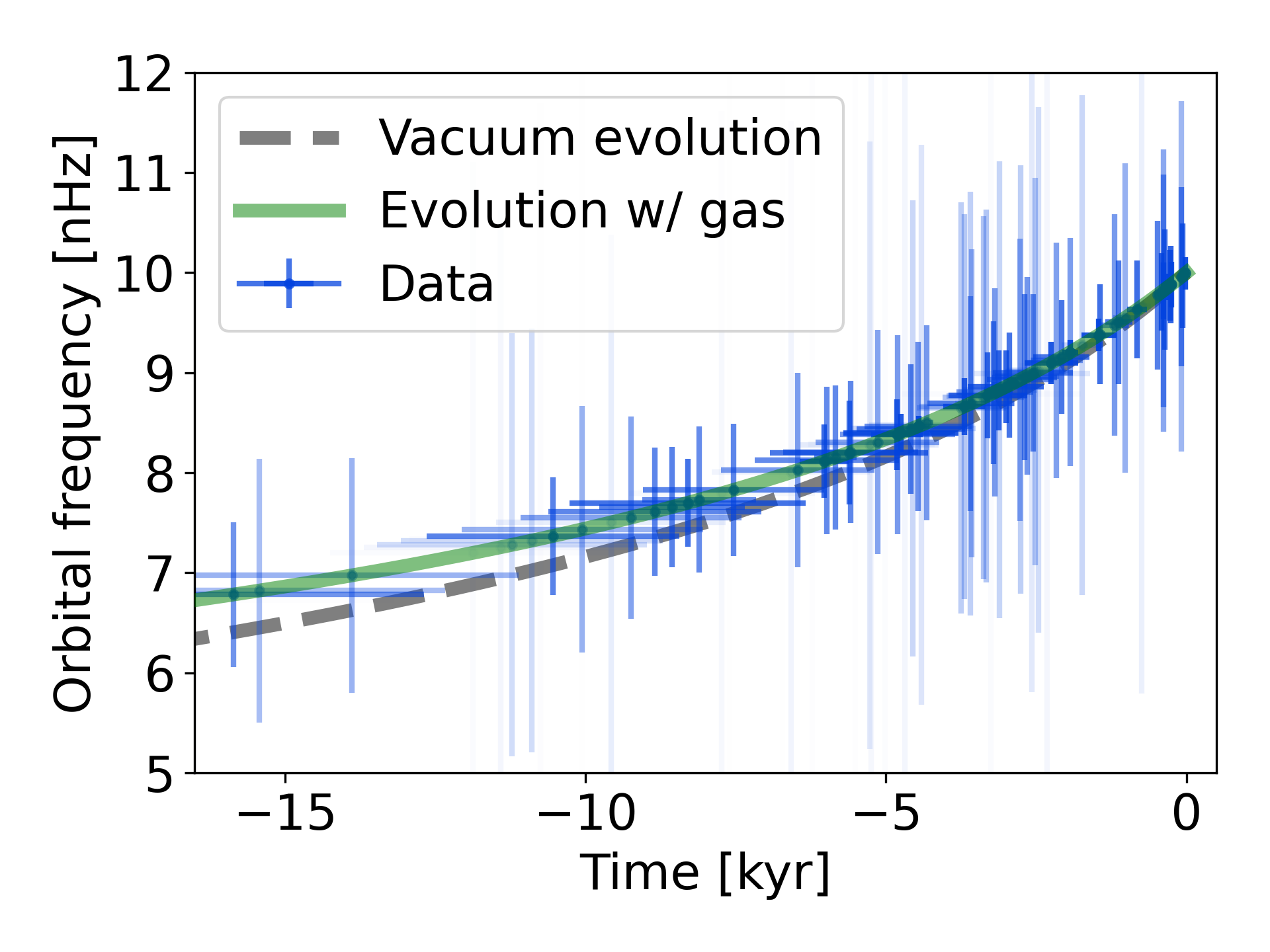}
    \caption{Example frequency evolution track of a $10^9 \, M_{\odot}$ chirp-mass equal-mass circular binary with an Earth-term orbital frequency of 10 nHz in vacuum (dashed line) and in the presence of gas with $f_{\rm Edd} = 100$ (solid line). Note that for equal-mass circular binaries, \citet{Siwek2023} predicts positive torques, which results in a slowed down inspiral visible here. Also shown are simulated frequency measurements from the pulsar terms of 116 IPTA pulsars, with more precise measurements shaded darker. The error bars show 1$\sigma$ uncertainties on the measured pulsar term frequency from the Fisher formalism (see Eq.~(\ref{eq:fisher})) and on the corresponding time due to pulsar distance uncertainty (see Eq.~(\ref{eq:psr_time})).}
    \label{fig:f_vs_t_example}
\end{figure}

These $t_i$ times also have associated errors. These are dominated by the low precision we have on the pulsar distances, so we ignore errors on $\cos \beta_i$. This is well-justified because we expect to localize these sources at least within a few tens of degrees by the time we can hope to precisely track their frequency evolution \citep{PolinaHostGalaxies2024, PolinaFirstCWII2025}. Note that this assumption could change in the future if we get better pulsar distance measurements, in which case one might even be able to phase-connect the signal \citep{Kato_psr_distance_cw}, so instead of just measuring frequency vs.~time, the pulsar terms could help map out GW phase vs.~time. Figure \ref{fig:f_vs_t_example} shows an example of what these frequency vs.~time measurements could look like: it shows frequencies and times measurable in 116 pulsar terms in an IPTA-like array with a 30-year baseline. We can see how the pulsar terms sample different times in the past and map out the frequency evolution of the source. Also shown is the frequency evolution of the simulated $10^9 \, M_{\odot}$ chirp-mass equal-mass circular binary with an Earth-term orbital frequency of 10 nHz in vacuum (dashed line) and in the presence of a gaseous environment. As we will see in Section~\ref{ssec:gas_drag}, equal-mass circular systems experience positive gas torques (outspiral), which results in the slower frequency evolution we see in the plot. The same system is investigated in more detail in Section~\ref{sec:results}.

We stress that the approach outlined above and thus our results below are approximate, as they rely on the Fisher matrix frequency errors, which require posteriors well-approximated by Gaussian distributions to remain accurate. In principle, this approximation can break at low SNR, where posteriors are not expected to be Gaussian. In particular, it is possible that while the overall SNR of the signal is high, its SNR in a particular pulsar is low, and thus the frequency errors for the corresponding pulsar term are inaccurate. However, such low-SNR pulsars are not expected to contribute significantly to the overall frequency evolution measurement. In addition, our treatment above ignores uncertainty in the source sky location, which would introduce additional uncertainty in the $t_i$ (see Eq.~(\ref{eq:psr_time})). Finally, we assume that uncertainties in the different pulsar term frequencies are independent. These approximations motivate future injection studies with a full Bayesian pipeline. However, to validate our results, we did carry out a small number of tests, where a binary with GW-only evolution was analysed both with the Fisher approach and a full Bayesian pipeline. By comparing the precision of the recovered chirp mass, we found that the Fisher approximation is accurate within 20-30\% for the relatively high SNRs considered in this paper.

\subsection{Gas drag model} \label{ssec:gas_drag}

To model the secular orbital evolution of an SMBHB embedded in a circumbinary disk (CBD), we can write the time derivative of the semi-major axis as:
\begin{equation}
    \dot{a}_{\rm tot} = \dot{a}_{\rm GW} + \dot{a}_{\rm gas},
    \label{eq:adot}
\end{equation}
where $\dot{a}_{\rm GW}$ prescribes the orbital decay due to GW emission, while the second term gives the effect of the gas environment. We numerically integrate Eq.~(\ref{eq:adot}) to get the orbital evolution, $a(t)$, which then through Kepler's 3rd law gives the evolution of the orbital frequency $f_{\rm orb}(t)$, and the gravitational-wave frequency evolution for a circular orbit $f_{\rm GW} (t) = 2f_{\rm orb}(t)$. The contribution of the gas environment is described based on \citet{Siwek2023} as:
\begin{equation}
    \dot{a}_{\rm gas} = \chi_{\rm CBD} (q,e=0) \frac{\dot M}{M_{b}} a = \chi_{\rm CBD} (q,e=0) \frac{f_{\rm Edd} L_{\rm Edd}}{\epsilon M_b c^2} a,
    \label{eq:adot_gas}
\end{equation}

where $\chi_{\rm CBD} (q,e)$ are the numerically determined values of $\dot{a}/a$ given in the left panel of Figure 2 of \citet{Siwek2023}, which we interpolate to arbitrary values of the binary mass ratio ($q$), while setting the orbital eccentricity $e=0$.\footnote{Note that the assumption of circular binaries is self-consistent given that $\dot{e}=0$ for circular binaries in \citet{Siwek2023}. However, extending our formalism to eccentric binaries is well-motivated and will be explored in future work.} To convert the dimensionless value of $\chi_{\rm CBD} (q,e)$ to an orbital evolution rate, an accretion rate onto the binary has to be inferred in order to achieve a physical scale for the influence of the disk on the binary. Here, we parameterize the accretion rate using the Eddington-limited accretion model, where $f_{\rm Edd}$ is the Eddington fraction, $L_{\rm Edd}$ is the Eddington luminosity, $c$ is the speed of light, and $M_b$ is the total mass of the binary system. Note that since $L_{\rm Edd}\propto M_b$, the gas-driven semi-major axis evolution is independent of $M_b$. We set the radiative efficiency to $\epsilon=0.1$, disk aspect ratio to $h_0=0.1$, and viscosity parameter to $\alpha=0.1$.\footnote{Setting different values of these parameter is degenerate with changing the Eddington fraction, see Eq.~(\ref{eq:fdb_fedd}) in Appendix \ref{app:disk_mass}.} Note that $f_{\rm Edd}$ in our model acts as a convenient scaling variable, which absorbs changes in all other model parameters, which we fix. 

For instance, lowering the disk scale height or viscosity in our parameterization would lead to an increase in the disk to binary mass ratio (cf.~Eq.~(\ref{eq:fdb_fedd}) in Appendix~\ref{app:disk_mass}), implying that gas-driven orbital evolution is stronger. Moreover, recent simulations (e.g., \citealt{Tiede2025}) show a buildup of gas at the inner edge of the cavity in thinner disks, which suggests that such systems further enhance gas torques on the binary. However, these simulations also show that accretion in such systems is suppressed for long timescales, as gas builds up at the cavity edge. Thus, increasing $f_{\rm Edd}$ -- i.e., the inflow rate of gas through the disk -- to and above the Eddington limit as a scaling parameter for the disk mass does not necessarily imply unphysically fast mass growth of the binary, as the inflow rate through the disk and the net accretion rate of the binary may decouple for long periods of time. On the other hand, we note that the orbital evolution of binaries embedded in thinner disks may be qualitatively different (e.g., \citealt{Tiede2020}), and may result in different signs and magnitudes of $\chi_{\rm CBD}$, an effect not considered in this study.
For more details, see Appendix \ref{app:disk_mass}, where we relate $f_{\rm Edd}$ and other parameters to the disk mass, which physically governs the strength of gas torques.

$\chi_{\rm CBD} (q,e)$ implicitly measures the interaction between the binary and the CBD through both gravitational torques and accretion torques. The instantaneous numerical value of $\chi_{\rm CBD} (q,e)$ varies significantly on binary orbital and disk precession timescales. Therefore, the final value of $\chi_{\rm CBD} (q,e)$ is obtained by running long numerical simulations and averaging the energy and angular momentum rate of change over several thousand binary orbits, yielding the net effect of the disk on the binary. 
Previous studies have indicated that the net torque acting between a binary and CBD
can cause orbital expansion \citep[``softening";][]{Miranda2016, Moody2019, Munoz2019, Duffell2020, Munoz2020, DorazioDuffell2021}, 
and the orbital evolution rates we take from \cite{Siwek2023} generalize this to a wide range of eccentricities and mass ratios.
In this work, we only consider circular binaries, for which the semi-major axis is found to shrink only for binaries with small mass ratios, while binaries with $q \gtrsim 0.25$ expand due to CBD interaction.

We note that in the case of binaries close to merger, the steady-state and scale-free assumptions made in the simulations of \citet{Siwek2023} may break down, as the binary may merge on a timescale shorter than that over which the orbital evolution rates are averaged to produce the values in Table 2 of \citet{Siwek2023}. In such cases, numerical simulations with a specific scale (such as binary mass, separation and disk mass) would be required to predict the gas-driven evolution mode accurately.

\subsection{Generic model of environmental effects} \label{ssec:generic_env}

A simple model of environmental effects in general inspired by \citet{speri_emri_environmental_effects} is where the time derivative of the semi-major axis is modeled as a power-law:
\begin{equation}
    \dot{a}_{\rm env} = -A_{\rm env} \left( \frac{a}{a_0} \right)^{\gamma_{\rm env}},
\label{eq:generic_adot}
\end{equation}
where $a_0$ is a reference value of the semi-major axis, which we choose to be the semi-major axis the binary has at the time corresponding to emitting GWs at the Earth-term frequency, $f_{\rm E}$. This is equivalent to a model where $\dot{f}\propto f^{\alpha_{\rm env}}$, with $\alpha_{\rm env}=5/3-2/3\gamma_{\rm env}$. We know that many hardening mechanisms can be described as power-laws. GW-driven hardening corresponds to $\alpha_{\rm env}=11/3$, $\gamma_{\rm env}=-3$; three-body interactions can produce $\alpha_{\rm env}=1/3$, $\gamma_{\rm env}=2$ \citep{Quinlan1996,Sampson2015}; and gas-disk interactions can produce $\alpha_{\rm env} \in [1,2]$, $\gamma_{\rm env} \in [-1/2,1]$ depending on the disk model used \citep{Ivanov1999,Haiman2009,SarahPTAReview,Siwek2023}. In particular, the model described in detail above gives $\gamma_{\rm env}=1$. Dynamical friction from a dark matter halo can also be approximated as a power-law  \citep{pippa_lisa_environmental_effects, dm_spike_TiruvaskarGordon2026}.

\subsection{Frequency-evolution fit}
\label{ssec:freq_evol_fit}

Using the Fisher-matrix errors on the measured frequencies from Section \ref{ssec:fisher}, along with a model for the frequency evolution from Section \ref{ssec:gas_drag} or \ref{ssec:generic_env} we can estimate how well we could measure parameters describing the binary and the environment. To do so, we need to fit a model of frequency vs.~time, $f(t, \theta)$, to measured $(f_i, t_i)$ values from each pulsar term and $(f_{\rm E}, t=0)$ from the Earth term, where $\theta$ are parameters of the model. Since each data point has associated errors in both parameters, $(\sigma_{f_i}, \sigma_{t_i})$, we carry out a simple Bayesian fit, where frequency errors are explicitly accounted for in the likelihood, while time errors are marginalised over numerically. For each data point, the time coordinate is replaced by a set of $N_{\rm marg}$ quadrature points
$t_{ij}=t_i+\sigma_{t_i}\Phi^{-1}[(2j-1)/(2N_{\rm marg})]$,
where $\Phi^{-1}$ is the inverse standard-normal CDF. The resulting log-likelihood is approximated as
\begin{equation}
\log \mathcal{L}(\theta)
\simeq
\sum_i
\log\left[
    \frac{1}{N_{\rm marg}}
    \sum_{j=1}^{N_{\rm marg}}
    \exp\left(
        -\frac{\left[f_i-f(t_{ij};\theta)\right]^2}
        {2\sigma_{f,i}^2}
    \right)
\right].
\end{equation}
This effectively averages the likelihood contribution over the Gaussian uncertainty distribution in $t_i$, avoiding the need to sample one additional latent time variable per observation. We then sample this low-dimensional likelihood with \texttt{dynesty} \citep{dynesty} to get posteriors on the model parameters describing the frequency evolution.

\section{Results} \label{sec:results}

To showcase the ability of the Fisher formalism introduced in Section \ref{ssec:fisher} to predict the detectability of environmental effects, we analyse a set of simulated datasets. These are made to resemble a future IPTA dataset with 116 pulsars and a 30-year observing baseline. Their creation follows the methodology of \citet{Astro4Cast2021} and \citet{PolinaFirstCWII2025}. We start with 45 pulsars from the NANOGrav 12.5-year dataset \citep{nanograv_12p5yr_data}, for which we use the measured noise parameters and observing epochs. We add 23 pulsars from the NANOGrav 15-year dataset \citep{nanograv_15yr_data}, 14 pulsars from PPTA DR3 \citep{ppta_dr3_data}, 3 pulsars from EPTA+InPTA DR2new+ \citep{epta_dr2_data}, and 31 pulsars from MPTA DR1 \citep{MPTA_4p5yr_data}. We extend all these observing epochs using the existing cadence to achieve a largest baseline of 30 years. Importantly, we use the real pulsar distances, which have a mean value of 1.6 kpc.

We analyse this base dataset with different signals injected to answer three key questions:
\begin{enumerate}
    \item How well can the parameters of an environment be measured? (Section~\ref{ssec:inference})
    \item What limits can we put on environments in the absence of a clear detection? (Section~\ref{ssec:limits})
    \item Can the inferred binary parameters be biased if we neglect to model environmental effects? (Section~\ref{ssec:bias})
\end{enumerate}

\subsection{Inference of environment parameters} \label{ssec:inference}

To test parameter inference of environmental effects we analyse datasets where both the simulated signal and the model used to recover it have environmental effects included. We simulate an equal-mass circular binary with Earth-term GW frequency of 20 nHz, a chirp mass of $10^9 \ M_{\odot}$, and ${\rm SNR}\simeq12$. The source sky location is chosen to be on the equatorial plane at right ascension of 0 h. 

First we consider a scenario where this binary is in a CBD environment. Our CBD model introduces two additional free parameters (see Eq.~(\ref{eq:adot_gas})): i) $f_{\rm Edd}$, which effectively controls the strength of the effect; and ii) $q$, which affects both the strength and the sign of the gas torques through the simulations from \citet{Siwek2023}. We simulate the signal with $f_{\rm Edd}=100$ to ensure a strong effect. This corresponds to a disk mass of about 1\% of the binary mass at the Earth term frequency (cf.~Eq.~(\ref{eq:fdb_fedd})). Note that since $f_{\rm Edd}$ acts as an overall scaling in our model, such a high value of it does not need to correspond to super-Eddington accretion. For example, the same disk mass torquing the binary can be achieved via a lower $f_{\rm Edd}$ value if other disk parameters are adjusted, see Eq.~(\ref{eq:fdb_fedd}) in Appendix \ref{app:disk_mass}. We also set $q=1$, which means that based on \citet{Siwek2023}, torques from the CBD are positive ($\dot{a}_{\rm gas}>0$), and will slow down the inspiral. In our analysis, we use uniform priors on all sampled parameters with ranges $f_{\rm E} \in [15, 25] \ \rm{n Hz}$, $\mathcal{M}\in[0,10^{10}] \ M_{\odot}$, $q \in [0.1,1.0]$, $f_{\rm Edd}\in[0,1000]$.

\begin{figure}[htbp!]
    \centering
    \includegraphics[width = 0.75\columnwidth]{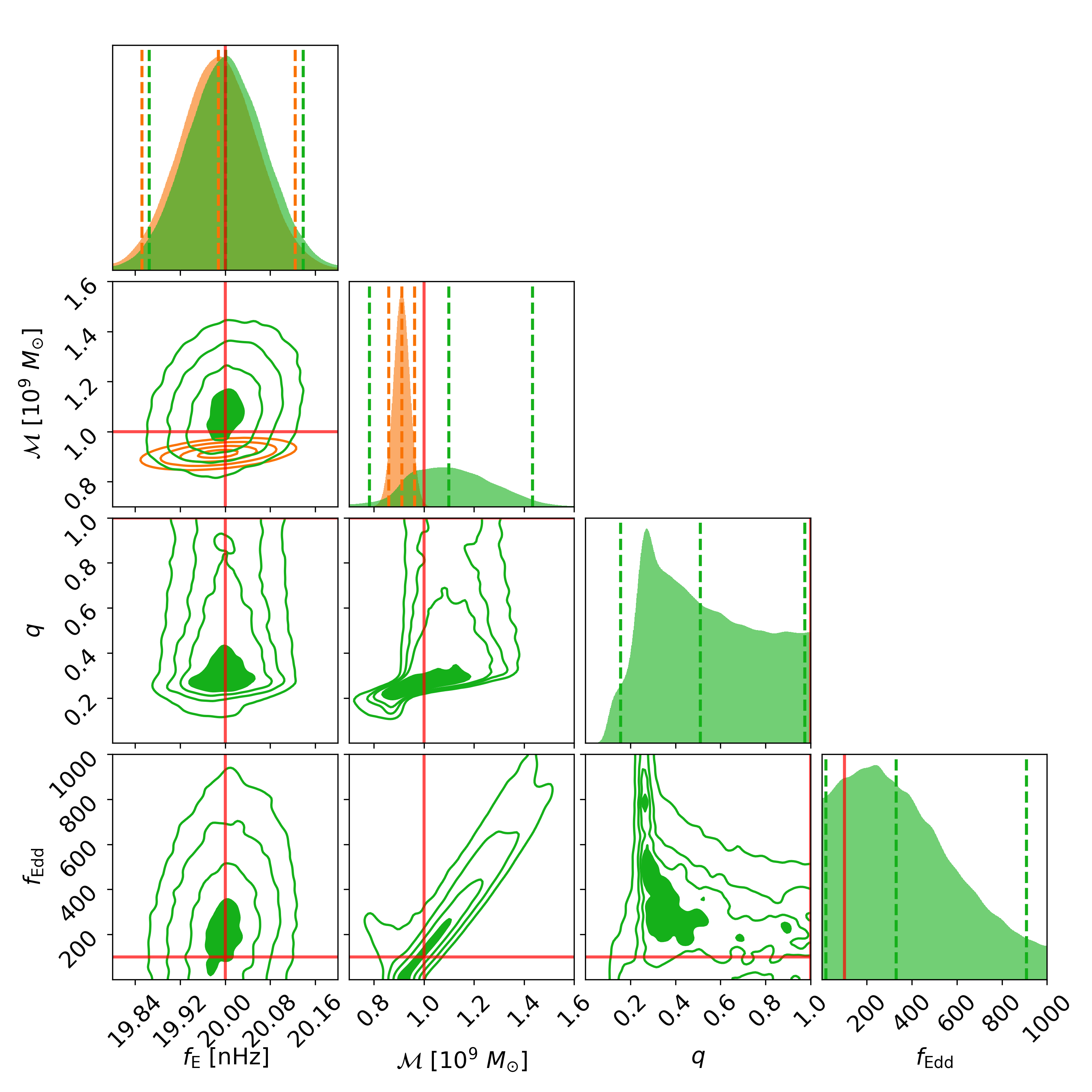}
    \caption{Inference on an SMBHB in a gaseous environment ($\mathcal{M}=10^9 \, M_{\odot}$, $f_{\rm E}=20$ nHz, $q=1$, $f_{\rm Edd}=100$) with (green) and without (orange) environmental effects included in the model. We can see that the recovered chirp mass is biased if the environment is not modeled. Modeling the environment avoids the bias, but the uncertainty on the chirp mass is much greater due to the degeneracy between the environment and GW evolution. This is because an overestimated chirp mass, with increasing $f_{\rm Edd}$ and $q\gtrapprox0.25$ can replicate a similar frequency evolution. Note that there is also some support for an underestimated chirp mass, with  $q\lessapprox0.25$. This transition is due to the fact that for circular binaries, the gas drag causes inspiral for $q\lessapprox0.25$ and outspiral otherwise (see Figure~2 in \cite{Siwek2023}), which can partially counteract the effect of an under or overestimated chirp mass, respectively.}
    \label{fig:gas_env_corner}
\end{figure}

Figure~\ref{fig:gas_env_corner} shows the posterior distributions of the binary and environment parameters (green). We also show results with environmental effects not modeled as a reference (orange). Red lines mark the true simulated values of parameters. We can see that using the vacuum frequency evolution can result in a biased recovery of the chirp mass (this will be further explored in Sec.~\ref{ssec:bias}). However, when environmental effects are modeled, the GW frequency and chirp mass are well-measured, while the mass ratio and Eddington fraction are only weakly constrained. This is expected given that the environmental effects are subdominant compared to the GW-driven evolution in this scenario. We can also notice some interesting correlation pattern between $\mathcal{M}$ and $f_{\rm Edd}$. This indicates that a higher chirp mass can fit the data similarly well if $f_{\rm Edd}$ is also increased. This is because the positive torques of the environment can practically cancel out GW-driven inspiral. Interestingly, a lower chirp mass is also permitted with a higher $f_{\rm Edd}$, resulting in a ``V'' shaped posterior. Also note the transition in $q$, where lower chirp mass corresponds to $q\lesssim0.25$, while a higher chirp mass corresponds to $q\gtrsim0.25$. This is due to the fact that the CBD drag switches sign around $q\simeq0.25$ (see Fig.~2 in \citealt{Siwek2023}), so for $q\lesssim0.25$ the negative CBD torques require a decreased chirp mass to give a similar amount of total frequency evolution.

\begin{figure}[htbp!]
    \centering
    \includegraphics[width = 0.6\columnwidth]{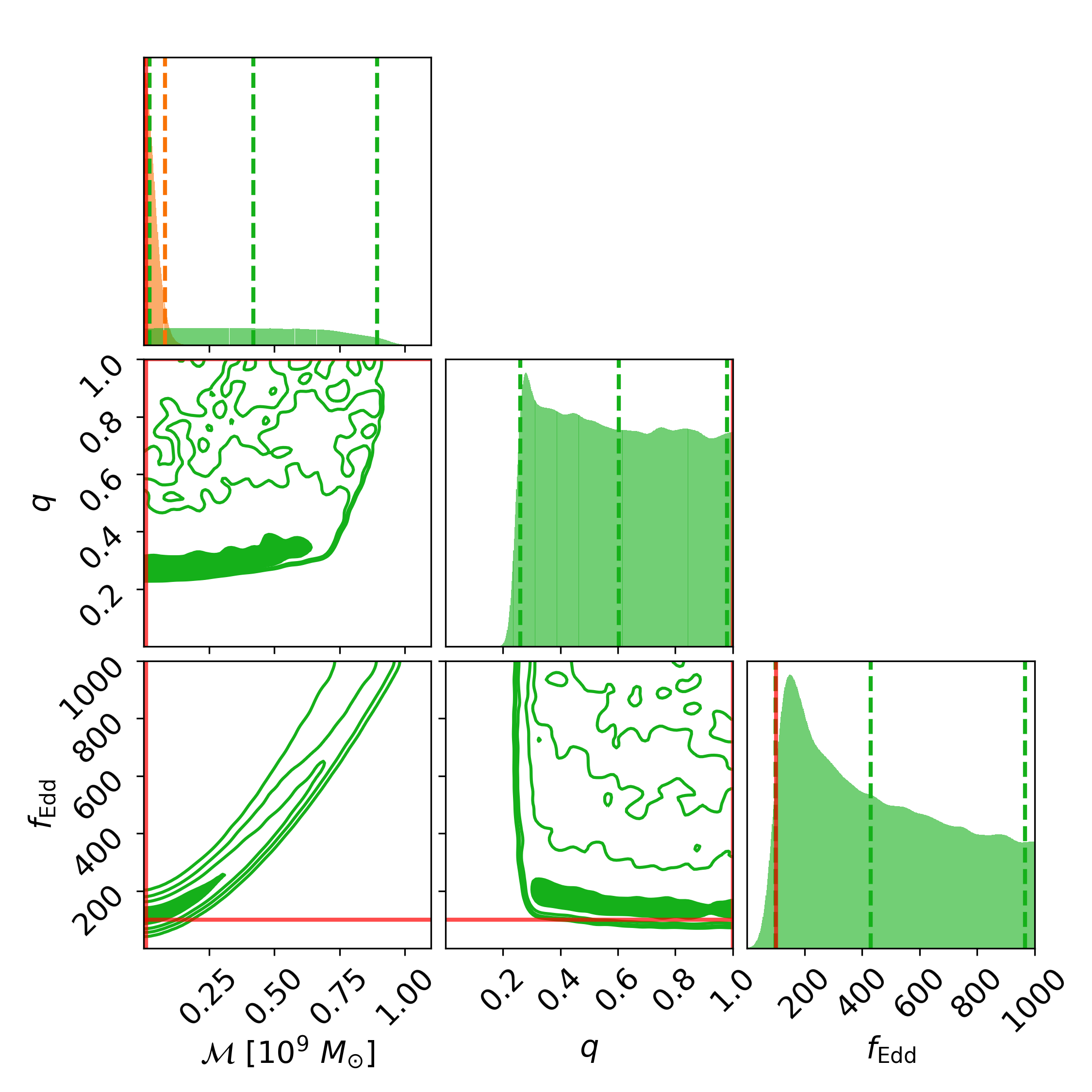}
    \caption{Same as Fig.~\ref{fig:gas_env_corner} (without the panel on $f_{\rm E}$), but with a lower chirp mass of $\mathcal{M}=10^7 \ M_{\odot}$, which results in negligible GW-driven inspiral, so the system ends up outspiraling due to the environmental effects. This can only be explained with an environment, so in this case $f_{\rm Edd}=0$ is clearly excluded and the environment is detected. However, covariance with GW-driven inspiral still means that $f_{\rm Edd}$ cannot be measured precisely, because very high values of $f_{\rm Edd}$ can be compensated for by a higher chirp mass.}
    \label{fig:outspiral_gas_env_corner}
\end{figure}

The situation is slightly different if the same environment is seen with a lower-mass binary (or lower frequency), where the GW-driven evolution is significantly slower. In this case, the evolution driven by the environment can dominate the overall evolution, leading to an outspiraling binary. Figure~\ref{fig:outspiral_gas_env_corner} illustrates this by showing the results of a similar analysis on a $\mathcal{M} = 10^7 \ M_{\odot}$ binary. We can see that since outspiraling frequency evolution can only be explained by an environment, the presence of an environment is clearly detected, as indicated by the posterior strongly disfavoring $f_{\rm Edd}=0$ (green posteriors in bottom row of Fig.~\ref{fig:outspiral_gas_env_corner}). However, the precise value of $f_{\rm Edd}$ is still not measurable due to the covariance with chirp mass, such that an overestimated $f_{\rm Edd}$ can be compensated for by a larger $\mathcal{M}$. We can also see that the mass ratio is constrained to $q\gtrsim0.25$ corresponding to the transition in the gas drag torque sign. It is important to note that $f_{\rm Edd}$ and $q$ cannot be measured separately beyond the sign of the torque, as they both just scale $\dot{a}$, but do not influence its scaling with $a$.

\begin{figure}[htbp!]
    \centering
    \includegraphics[width = 0.75\columnwidth]{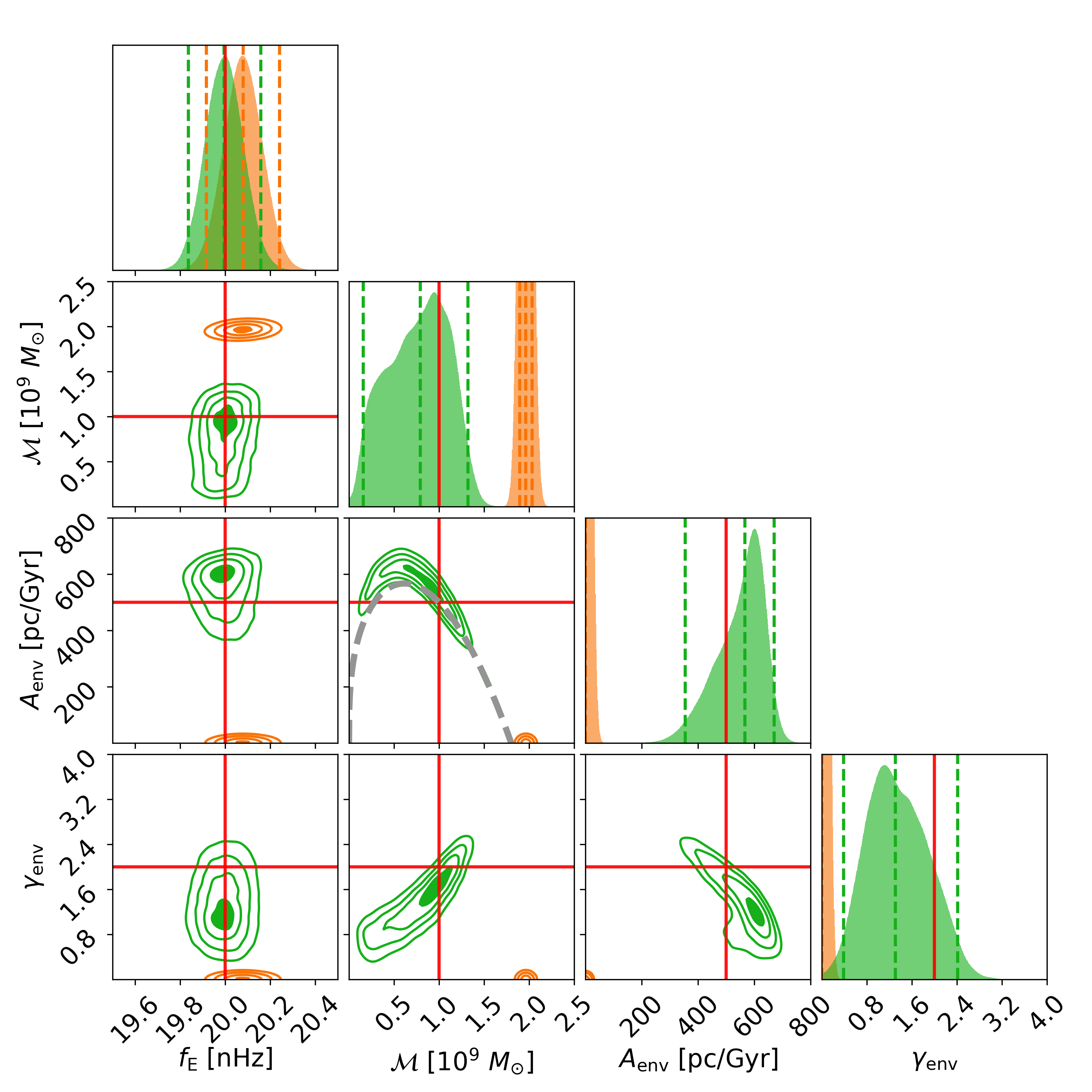}
    \caption{Inference on an SMBHB in a generic environment ($\mathcal{M}=10^9 \, M_{\odot}$, $f_{\rm E}=20$ nHz, $\gamma_{\rm env}=2$ and $A_{\rm env} = 500$ pc/Gyr) with (green) and without (orange) environmental effects included in the model. The exclusion of environmental effects results in a strong bias in the recovered chirp mass, while correct modeling of the environment allows for an accurate recovery of both the chirp mass and the parameters of the environment.}
    \label{fig:powerlaw_env_corner}
\end{figure}

We also test a scenario where the environmental effect is prescribed by our generic power-law description from Eq.~(\ref{eq:generic_adot}). Here we simulate the same 20 nHz, $10^9 \ M_{\odot}$ chirp mass binary as above, but with a generic environment with $\gamma_{\rm env}=2$ and $A_{\rm env} = 500$ pc/Gyr. We chose this example as it makes the frequency dependence more different from GW-driven evolution ($\gamma_{\rm env}=-3$) than the gas-drag model above ($\gamma_{\rm env}=1$). This also happens to be the expected dependence for a stellar scattering driven evolution \citep{Quinlan1996, Sampson2015}. We set uniform priors on all sampled parameters with ranges $f_{\rm E} \in [15, 25] \ \rm{n Hz}$, $\mathcal{M}\in[0,10^{10}] \ M_{\odot}$, $A_{\rm env} \in [0,2] \ \rm{kpc/Gyr}$, $\gamma_{\rm env}\in[-8,8]$. Figure~\ref{fig:powerlaw_env_corner} shows our results on this system with both a model including environmental effects (green) and one without (orange). In this case, the extra parameters we sample are the two free parameters from Eq.~(\ref{eq:generic_adot}), which determine the strength ($A_{\rm env}$) and frequency dependence ($\gamma_{\rm env}$) of the environmental effect. We can see that in this particular example, the environmental effect is strong enough to be clearly detected and distinguished from GW-only evolution. This is not only due to the strength of the environmental effects in this example, but also the large difference in $\gamma$, which breaks the degeneracy between GW-driven and environment-driven evolutions. We can also see that both $A_{\rm env}$ and $\gamma_{\rm env}$ can be measured accurately.

It is interesting to note the non-linear covariance structure visible in Fig.~\ref{fig:powerlaw_env_corner} between the measured $\mathcal{M}$ and $A_{\rm env}$ value. This is due to the fact that the environment is parametrized in terms of $\dot{a}$, but what is measured is $\dot{f}$, and the conversion is dependent on the total mass of the system through Kepler's third law, which in turn is related to the chirp mass. So while physically the environmental effect is mass-independent, in our parametrization, it does have a slight mass dependence. This is illustrated in Fig.~\ref{fig:powerlaw_env_corner} with the gray dashed line, which corresponds to $A_{\rm env}$ values resulting in the same total frequency derivative at the Earth term at different $\mathcal{M}$ values, taking into account the above mentioned weak chirp-mass-dependence of $\dot{f}_{\rm env}$, which results in the otherwise linear inverse relationship to curve down at small $\mathcal{M}$ values.

\subsection{Constraining the presence of environments} \label{ssec:limits}

We also investigate the extent to which environmental effects can be constrained in the absence of a definitive identification. For this purpose we simulate a binary system with no environmental effects, but analyze it with the model that includes environmental effects described by our generic model from Eq.~(\ref{eq:generic_adot}). Specifically, we look at the same equal-mass circular binary as in Sec.~\ref{ssec:inference} with $\mathcal{M} = 10^9 \, M_{\odot}$. As expected our results are consistent with $A_{\rm env} = 0$. 

\begin{figure}[htbp!]
    \centering
    \begin{minipage}[b]{0.6\columnwidth}
        \centering
        \includegraphics[width=\linewidth]{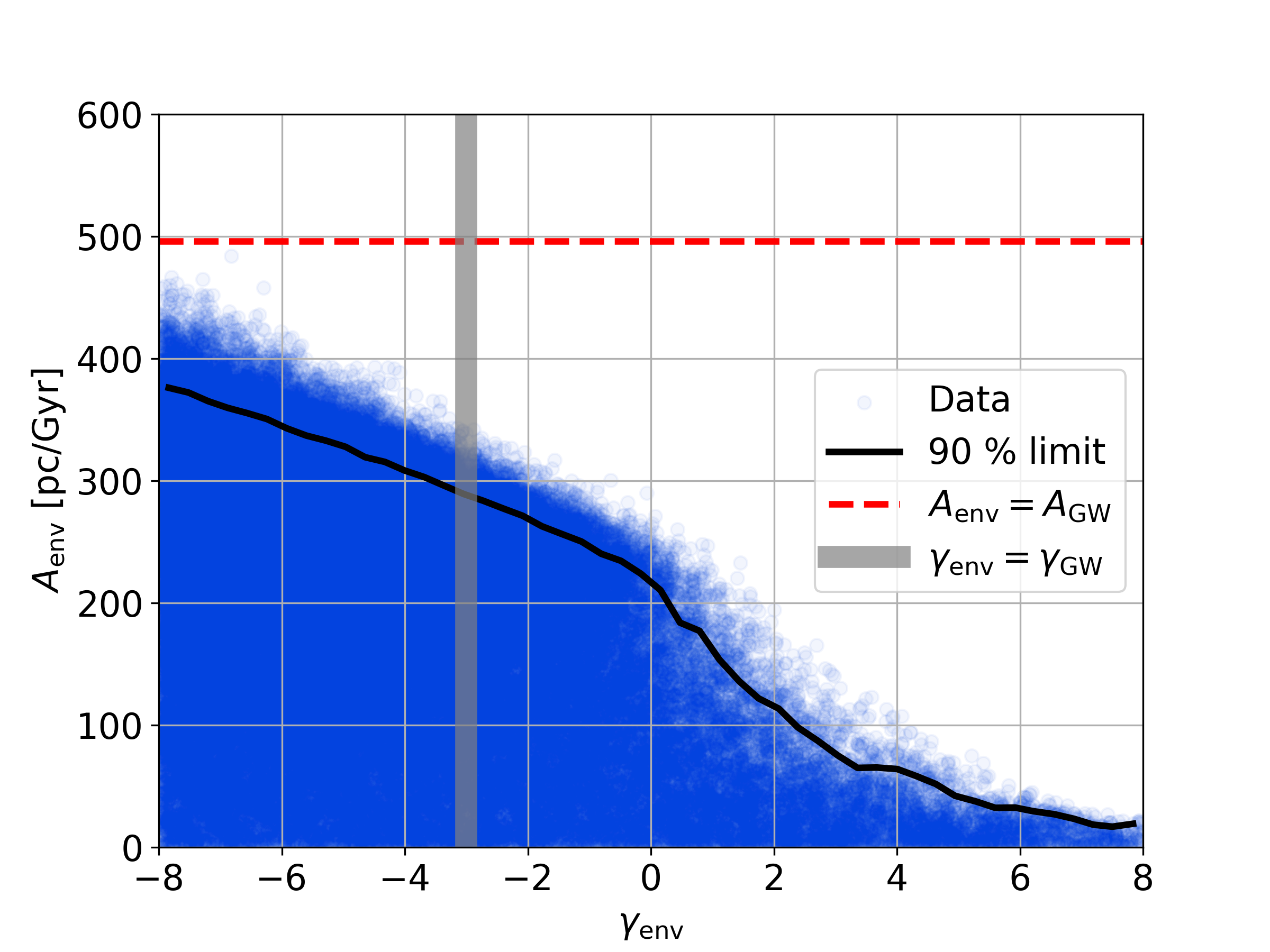}
        \label{fig:powerlaw_env_limit_a}
    \end{minipage}
    \hfill
    \begin{minipage}[b]{0.6\columnwidth}
        \centering
        \includegraphics[width=\linewidth]{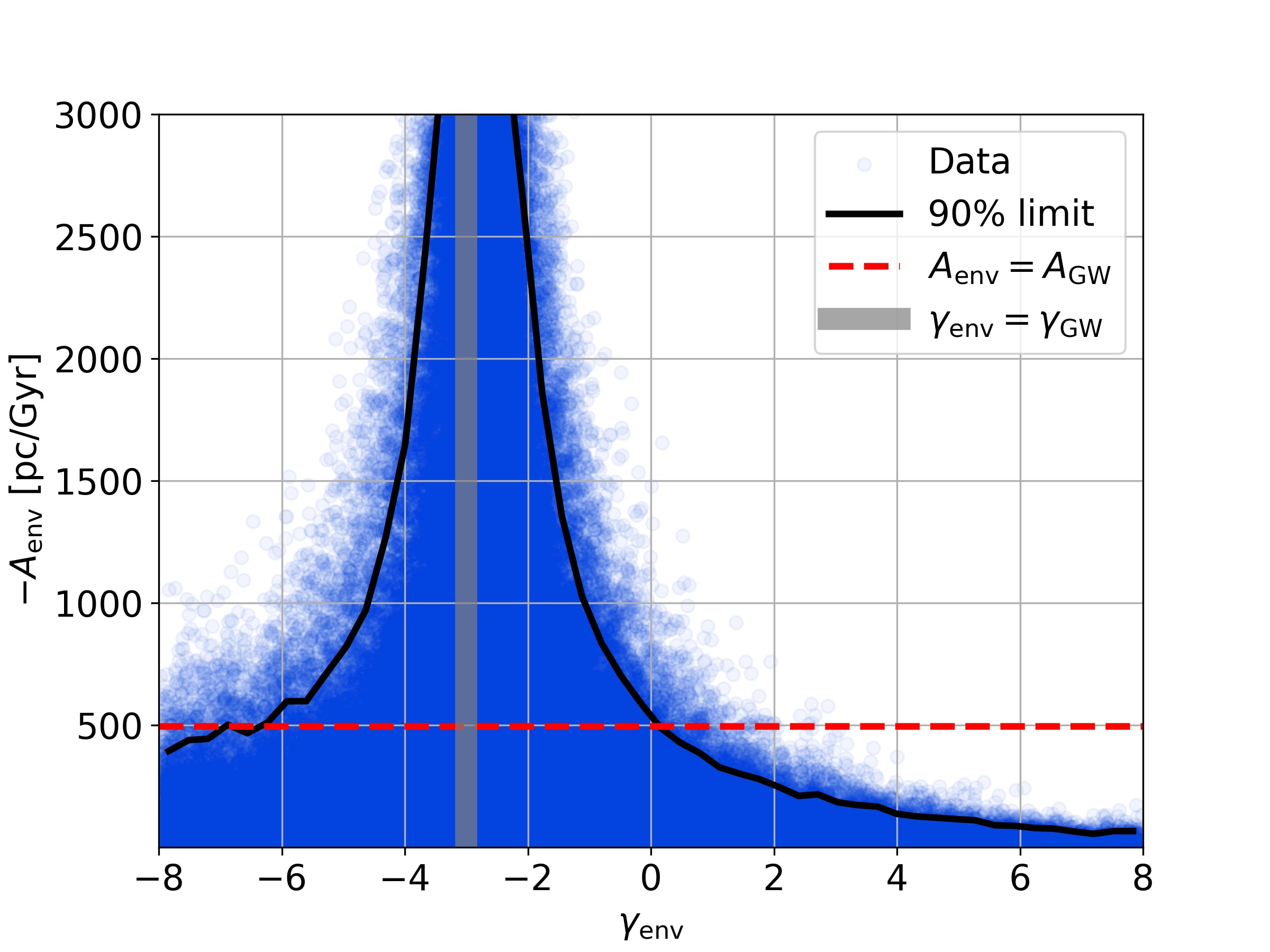}
        \label{fig:powerlaw_env_limit_b}
    \end{minipage}
    \caption{Example on constraining generic environments. We analysed a dataset with 30-year IPTA setup and an $\mathrm{SNR}\simeq12$, $\mathcal{M} = 10^9 \, M_{\odot}$, equal-mass circular binary with an Earth-term orbital frequency of 10 nHz assuming vacuum evolution, while allowing for a generic environment parametrised as in Eq.~(\ref{eq:generic_adot}). The top panel shows the upper limit we can put on the hardening amplitude ($A_{\rm env}$) as a function of $\gamma_{\rm env}$ for hardening ($A_{\rm env}>0$). The bottom panel shows the same for softening ($A_{\rm env}<0$) environments. Note that a softening environment with $\gamma_{\rm env} \approx -3$ cannot be constrained as this is degenerate with an increased chirp mass.}
    \label{fig:powerlaw_env_limit}
\end{figure}

Figure~\ref{fig:powerlaw_env_limit} shows the 90\% upper limit we can put on $A_{\rm env}$ at different $\gamma_{\rm env}$ values (black). Also shown are individual posterior samples (blue). The plot is separated into two panels corresponding to environments driving inspiral (top, $A_{\rm env}>0$) and outspiral (bottom, $A_{\rm env}<0$). We can see that the upper limits we get behave very differently for these two cases. For positive $A_{\rm env}$ values, our upper limits are of the same order of magnitude as the amplitude associated with GW-driven inspiral ($A_{\rm GW}$), with gradually stronger limits at higher $\gamma_{\rm env}$ values. This is expected, because the amplitude is referenced at the Earth-term frequency, so large positive values of $\gamma_{\rm env}$ lead to a stronger effect in the past where pulsar terms probe the frequency. For the case of negative $A_{\rm env}$ values, we can see that we cannot put any limit on them at $\gamma_{\rm env}=\gamma_{\rm GW}=-3$, because here GW-driven and environment-driven evolution is completely degenerate. The limits we can place on these outspiral-driven environments are thus strongly dependent on $\gamma_{\rm env}$, and get more and more constraining as we move away from $\gamma_{\rm env}=-3$. Also note that the overall scale of the plot indicates that we can typically only rule out quite strong environments with $A_{\rm env}$ values on the order of 10$^{1-3}$ of pc/Gyr. However, this is strongly dependent on the PTA setup and the parameters of the binary system considered. A more detailed parameter exploration is left for future work. 

\vspace{1 cm}
\subsection{Chirp mass stealth bias} \label{ssec:bias}

Even in scenarios where environmental effects cannot be clearly detected, they can lead to biased estimates of binary parameters, called stealth bias (see e.g.~\citet{Grace_LISA_stealth_bias2024} for an example in the case of LISA galactic binaries). We have seen in Section \ref{ssec:inference} that a gaseous environment needs to be fairly strong before it can be identified. Motivated by this, here we investigate how the presence of a gas environment can bias estimates of the chirp mass, even if the environment cannot clearly be identified. For this we analyze datasets with the same $10^9 \, M_{\odot}$ chirp-mass binary as before, but with different values of $f_{\rm Edd}$, controlling the strength of the environment. We analyze these without modeling environments, to see how much this unjustified simplification biases parameter recovery. 

\begin{figure}[htbp!]
    \centering
    \includegraphics[width = 0.6\columnwidth]{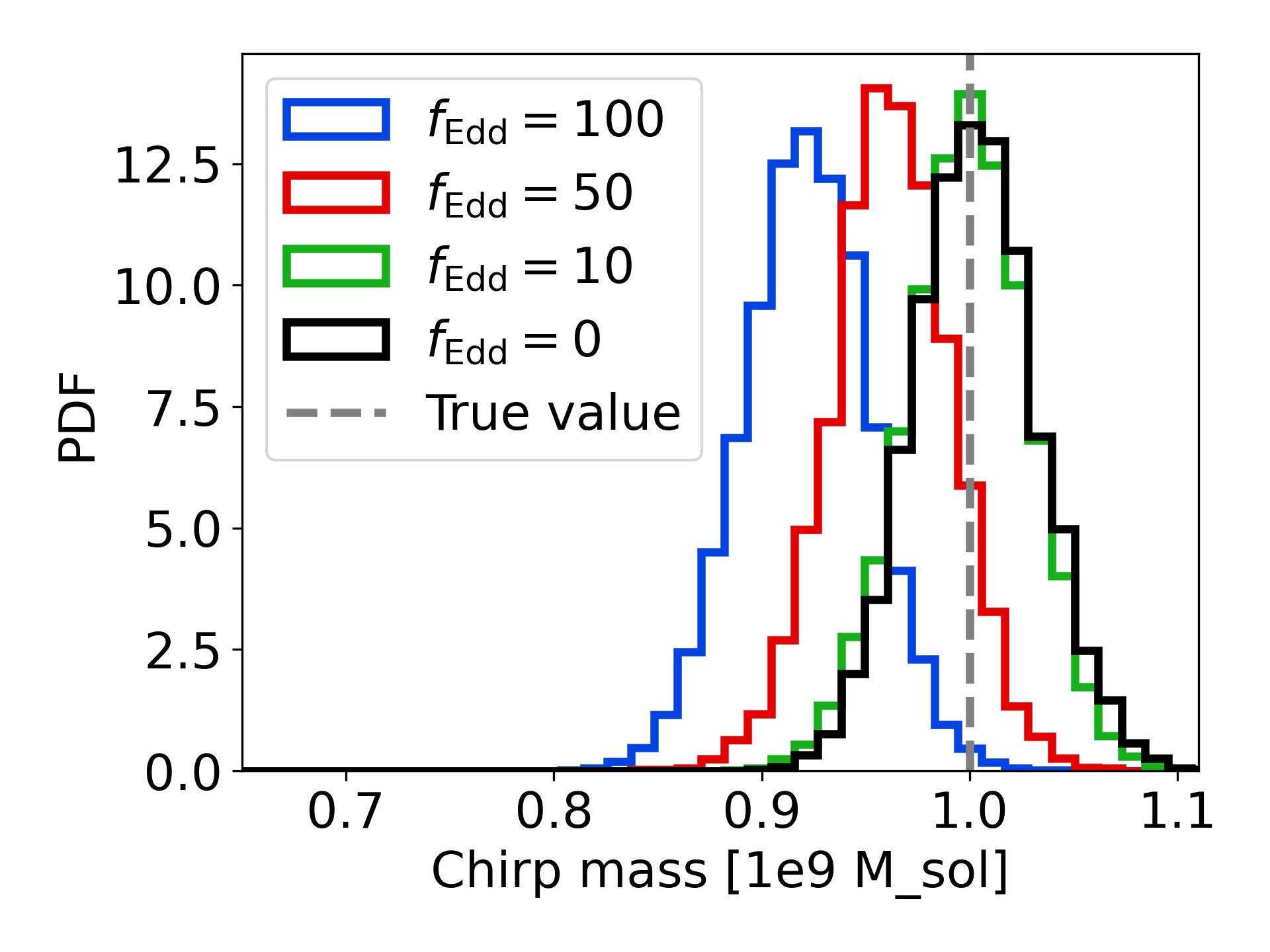}
    \caption{Illustration of chirp mass stealth bias with a 30-year IPTA setup. Shown are chirp mass posteriors for an SNR $\simeq12$, $10^9 \, M_{\odot}$ chirp-mass equal-mass circular binary with an Earth-term orbital frequency of 10 nHz assuming vacuum evolution. Various curves show results for different values of $f_{\rm Edd} = (100,50,10,0)$, which sets the strength of the gas torques. These approximately correspond to a disk mass of 1\%, 0.5\%, 0.1\%, and 0\% of the binary total mass at the Earth-term frequency, respectively. The chirp mass measurement becomes biased for large values of $f_{\rm Edd} \gtrapprox 50$.}
    \label{fig:chirp_mass_bias}
\end{figure}

Figure~\ref{fig:chirp_mass_bias} shows the posterior distributions of the chirp mass we recover for different values of $f_{\rm Edd}$. We can see that for moderately large values of the Eddington fraction ($f_{\rm Edd}\lesssim 10$) there is no significant bias. This corresponds to disk masses of less than 0.1\% of the binary total mass (cf.~Eq.~(\ref{eq:fdb_fedd})). However, for particularly strong environments ($f_{\rm Edd}\gtrsim 50$), the chirp mass can be significantly underestimated. This is because circular equal-mass binaries are expected to experience positive torques from the CBD, which leads to a slower inspiral that can be explained reasonably well without environmental effects and a lower chirp mass.

From our results, we can draw the preliminary conclusion that the ability of PTAs to accurately measure the chirp mass of an SMBHB will likely not be limited by environmental effects in the near future, as it would require an unexpectedly strong environment to cause significant bias. Note however, that this does not necessarily require super-Eddington accretion, as $f_{\rm Edd}$ captures additional parameter flexibility in our model (see Appendix~\ref{app:disk_mass}).  However, we should stress that this statement depends on several factors that should be explored in more detail in the future: i) at higher SNRs (i.e.~more sensitive PTA configurations) and correspondingly higher precision, the maximum tolerable bias will be lower, so less extreme environments could lead to significant bias; ii) we only tested a single Earth term frequency and chirp mass, but both the precision of chirp mass measurement and the overall frequency evolution depends strongly on these parameters -- a detailed parameter space exploration would be warranted in the future; iii) eccentricity can significantly impact the frequency evolution, and thus it is possible that a gaseous environment can have a stronger effect on an eccentric binary (see Section \ref{sec:conclusions} for more details).

\section{Conclusions and Future Work} \label{sec:conclusions}

Pulsar timing arrays will be able to detect supermassive black hole binaries in their early inspiral phase. In this paper, we explored how this will allow tracking the evolution of the binary over 10,000-year timescales through measurement of the pulsar terms. This could open the possibility of identifying effects other than gravitational wave emission that might affect the binary's evolution (e.g.~a gaseous environment, higher PN effects, eccentricity, dark matter, etc.). 

As a first exploration, we developed a simple Fisher-matrix formalism to predict the precision achievable by such a measurement for circular binaries with incoherent pulsar-term measurements (see caveats and future work below). We found that near future PTAs will likely be unable to clearly detect a gaseous environment (see Fig.~\ref{fig:gas_env_corner}). This is not only because of the weakness of this effect, but also because it is covariant with gravitational-wave driven evolution. We also find that the measured binary parameters, such as chirp mass, will not be significantly biased if the environmental effects are not modeled (see Fig.~\ref{fig:chirp_mass_bias}), unless the gaseous environment is unexpectedly strong. 

Our framework can also model generic power-law environments, which can approximately describe many physical scenarios, such as stellar scattering, gas torques, or dynamical friction in a dark matter halo. We demonstrate how a strong stellar scattering environment can be cleanly separated from gravitational wave only evolution (see Fig.~\ref{fig:powerlaw_env_corner}), thanks to its spectral index being substantially different from gravitational wave inspiral. We also show how constraints can be put on the presence of such a generic environment (see Fig.~\ref{fig:powerlaw_env_limit}), which is an approach that can yield potentially interesting physical constraints even if a detected signal is fully consistent with gravitational wave only inspiral. Similarly, such an analysis would be a powerful test of the expected GW-driven frequency evolution, and could be a great tool to confirm that a candidate signal is indeed coming from an SMBHB.

We make a few simplifying assumptions in this paper, which might affect our conclusions. As noted above, we only consider circular binaries. While gravitational wave emission is expected to circularize binaries, recent results in CBD modeling suggest that the presence of gas can lead to a stable eccentric orbit with an eccentricity around $e \simeq 0.3-0.5$ \citep{equilibrium_ecc_Zrake2021, DorazioDuffell2021, Siwek2023}. Eccentricity would not only spread the GW emission over several frequencies, but would also speed up the frequency evolution. Both of these would affect our ability to track the evolution and constrain environmental effects, and will be the subject of a future study. 

In this paper, we specifically focused on quantifying our ability to track the evolution only through frequency information. This is because frequency is the most easily measurable parameter of a sinusoidal signal, and pulsar term frequencies can be utilised even in the presence of significant pulsar distance errors. This essentially corresponds to what we would be capable if we detected an SMBHB now. In the future, however, we may have precise distance measurements for several pulsars, which could allow us to coherently detect the pulsar terms \citep{Deng_Finn2011,Taylor2026_localisation}. In this case, it would be possible to track the GW phase over time, $\Phi_{\rm GW} (t)$, instead of the frequency over time, $f_{\rm GW} (t) = \dot{\Phi}_{\rm GW} (t)/(2\pi)$. This in principle would allow much more precise tracking of the binary evolution \citep{Chiara_CW_evolution_tracking}, which will need to be investigated in more detail in the future. A related idea that was recently proposed is to track the phase evolution of binaries that merge at higher frequencies outside the PTA band \citep{orphan_PSR_term_Zheng2026,orphan_PSR_term_Criswell2026,orphan_PSR_term_Leclere2026}. For these, the Earth term may be observable by other experiments sensitive in the $\mu$Hz-mHz band, while the pulsar terms would reside in the PTA band, thus allowing long term tracking of the binary's evolution. 

Finally, in this paper, we used a relatively simple Fisher formalism. This is a good approximation at high signal-to-noise ratios, and it goes beyond simple counting of the accumulated phase or frequency evolution. However, future work will need to implement this in a full Bayesian analysis framework, and demonstrate how the frequency evolution can be tracked starting from the PTA time-of-arrival data.

We chose a binary with a chirp mass of $10^9 M_{\odot}$ and an Earth-term frequency of 20 nHz as our illustrative example, because it is broadly representative of the first SMBHB systems expected to be detected by PTAs (see e.g.~\citealt{RealisticDetection1, Emiko2025}). A more comprehensive exploration of the parameter space is nevertheless warranted, as we have only considered a small number of illustrative cases here. Understanding how these environmental effects depend on binary chirp mass and frequency, together with the expected astrophysical distribution of these systems, will ultimately determine the prospects for measuring them.

\begin{acknowledgments}
We thank Chiara Mingarelli for valuable feedback on the manuscript. B.B.~is supported by the UK’s Science and Technology Facilities Council, grant number UKRI5122. We also acknowledge the use of OpenAI's ChatGPT for support with code implementation, documentation, and manuscript refinement.
\end{acknowledgments}


\software{\texttt{dynesty} \citep{dynesty}, \texttt{enterprise} \citep{enterprise}, \texttt{numba} \citep{numba_paper, numba_zenodo}, \texttt{scipy} \citep{scipy}, \texttt{holodeck} \citep{holodeck_github}, \texttt{cw\_cbd} \citep{cwcbd_github}
          }

\appendix

\section{Disk mass fraction and Eddington ratio}
\label{app:disk_mass}

The strength of the gas drag model described in Section~\ref{ssec:gas_drag} is parametrized by the Eddington fraction $f_{\rm Edd}=\dot M/\dot M_{\rm Edd}$, where $\dot M$ is the mass accretion rate, and $\dot M_{\rm Edd}=L_{\rm Edd}/(\epsilon c^2)$ is the Eddington accretion rate, where $L_{\rm Edd}$ is the Eddington luminosity, and $\epsilon$ is the radiative efficiency. This is a convenient way to scale the model, however, mass accretion is not directly responsible for the torque experienced by the binary and the binary's resulting evolution. Instead, for a given disk configuration determined by the binary mass ratio and the orbital eccentricity, the torques depend on the mass of the disk's inner regions. We can parametrize this via the disk-to-binary mass fraction, $f_{\rm DB}\equiv M_{\rm disk}(2a,10a)/M_b$, where $M_{\rm disk}(R_1,R_2)$ is the mass in the circumbinary disk between two radii, $R_1$ and $R_2$, and $M_b=M_1+M_2$
and $a$ are the total binary mass and semimajor axis, respectively. We chose to define $f_{\rm DB}$ with a disk mass between $2 a$ and $10 a$ to focus on the innermost regions of the disk, but avoid the cavity. This parametrization highlights that what fundamentally determines the evolutionary impact of the gas on the binary is the amount of gas mass close to the binary orbit.  Here we relate these two parametrizations by calculating $M_{\rm disk}$ for a simple thin alpha disk model parametrized by $f_{\rm Edd}$. We adopt the surface-density
prescription of \citet{Munoz2020},
\begin{equation}
\Sigma(R)=f_{\rm cav}(R)\,\Sigma_0\left(\frac{R}{a}\right)^{-1/2}
\left[1-0.7\sqrt{\frac{a}{R}}\right],
\label{eq:disk_profile}
\end{equation}
where
\begin{equation}
\Sigma_0=\frac{\dot M}{3\pi\alpha h_0^2\Omega_b a^2}.    
\end{equation}

Here $h_0=H/R$ is the constant disk aspect ratio, $\alpha$ is the viscosity
parameter, and $\Omega_b=(GM_b/a^3)^{1/2}=2\pi f_{\rm orb}$ is the orbital angular
frequency. The normalization follows from the viscosity prescription
$\nu=\alpha h_0^2R^2\Omega_K$, with $\Sigma_0=\dot M/[3\pi\nu(a)]$.

Approximating the cavity factor as $f_{\rm cav}=1$ and writing $x=R/a$,
the mass between two radii is
\begin{equation}
M_{\rm disk}(R_1,R_2)=2\pi\int_{R_1}^{R_2}R\Sigma(R)\,dR
=\frac{2\dot M}{3\alpha h_0^2\Omega_b}
\left[\frac{2}{3}x^{3/2}-0.7x\right]_{R_1/a}^{R_2/a}.
\label{eq:disk_integral}
\end{equation}

Thus, for the adopted interval,
\begin{equation}
f_{\rm DB}\simeq\frac{9.0642}{\alpha h_0^2\Omega_b}\frac{\dot M}{M_b}.
\label{eq:fdb_mdot}
\end{equation}

The ratio of the accretion rate and the binary mass can be expressed as
\begin{equation}
\frac{\dot M}{M_b} = \frac{f_{\rm Edd} L_{\rm Edd}}{\epsilon c^2 M_{\rm b}} =\frac{f_{\rm Edd}}{\epsilon t_E},
\end{equation}
where
\begin{equation}
    t_E\equiv \frac{M_{\rm b} c^2}{L_{\rm Edd}} = \frac{\sigma_Tc}{4\pi Gm_p}\simeq4.50\times10^8\,\mathrm{yr},
\end{equation}
where $m_p$ and $\sigma_T$ denote the proton mass and Thomson cross section.
The conversion between the disk mass fraction and Eddington ratio is therefore
\begin{equation}
f_{\rm DB}\simeq\frac{9.0642}{\alpha h_0^2\Omega_b\epsilon t_E}\,f_{\rm Edd}
\simeq1.01\times10^{-4}\,f_{\rm Edd}
\left(\frac{\alpha}{0.1}\right)^{-1}
\left(\frac{h_0}{0.1}\right)^{-2}
\left(\frac{\epsilon}{0.1}\right)^{-1}
\left(\frac{f_{\rm orb}}{10^{-8}\,\mathrm{Hz}}\right)^{-1}.
\label{eq:fdb_fedd}
\end{equation}

We can see that for a given orbital frequency and disk parameters, $f_{\rm Edd}$ and $f_{\rm DB}$ fully determine one another. Note however that since we keep $\alpha=0.1$, $h_0=0.1$, and $\epsilon=0.1$ fixed in our disk model, $f_{\rm Edd}$ is the only free parameter that can change the disk mass, which is ultimately responsible for torquing the binary. This covariance highlights that high values of $f_{\rm Edd}$ in this model should not be read literally, as changes in other disk parameters could account for the same effect.

\bibliography{mybib}{}
\bibliographystyle{aasjournal}



\end{document}